\documentclass[fleqn,usenatbib]{mnras}

\usepackage{newtxtext,newtxmath}

\usepackage[T1]{fontenc}
\usepackage{ae,aecompl}

\usepackage{graphicx}	
\usepackage{amsmath}	
\usepackage{subfig}
\usepackage{xspace}
\usepackage{threeparttable}

\newcommand{\swift}{{\it Swift}\xspace}
\newcommand{\xmm}{\textit{XMM-Newton}\xspace}
\newcommand{\rosat}{\textit{ROSAT}\xspace}
\newcommand{\ergcm}{erg\,cm$^{-2}$\,s$^{-1}$}

\newcommand{\oergs}[1]{$10^{#1}$ erg s$^{-1}$}
\newcommand{\src}{{XMMU\,J050722.1$-$684758}\xspace}
\newcommand{\srce}{{MCSNR\,J0507$-$6847}\xspace}
\newcommand{\osrc}{{2MASS\,05072214$-$6847592}\xspace}

\newcommand{\SII}{[S\,{\sc ii}]}

\newcommand{\Halpha}{[H\,${\alpha}]$}

\usepackage{xcolor}

\title[ \src]{\src: Discovery of a new Be X-ray binary pulsar likely associated with the supernova remnant \srce}

\author[C. Maitra et al.]{C. Maitra,$^{1}$\thanks{E-mail: cmaitra@mpe.mpg.de}
F. Haberl,$^{1}$
P.~Maggi,$^{2}$
P.~J.~Kavanagh,$^{3}$
G.~Vasilopoulos,$^{4}$
\newauthor
M.~Sasaki,$^{5}$
M.~D.~Filipovi\'c,$^{6}$
A.~Udalski$^{7}$
\\
$^{1}$Max-Planck-Institut f{\"u}r extraterrestrische Physik, Gie{\ss}enbachstra{\ss}e, 85748 Garching, Germany\\
$^{2}$Universit\'e de Strasbourg, CNRS, Observatoire astronomique de Strasbourg, UMR 7550, F-67000 Strasbourg, France\\
$^{3}$School of Cosmic Physics, Dublin Institute for Advanced Studies, 31 Fitzwillam Place, Dublin 2, Ireland\\
$^{4}$Department of Astronomy, Yale University, PO Box 208101, New Haven, CT 06520-8101, USA\\
$^{5}$Remeis Observatory and ECAP, Universit{\"a}t Erlangen-N{\"u}rnberg, Sternwartstr. 7, 96049 Bamberg, Germany\\
$^{6}$Western Sydney University, Locked Bag 1797, Penrith South DC, NSW 2751, Australia\\
$^{7}$Astronomical Observatory, University of Warsaw, Aleje Ujazdowskie 4, 00-478 Warsaw, Poland\\
}

\date{Accepted XXX. Received YYY; in original form ZZZ}

\pubyear{2020}
\begin{document}
\label{firstpage}
\pagerange{\pageref{firstpage}--\pageref{lastpage}}
\maketitle

\begin{abstract}
We report the discovery of a new high mass X-ray binary pulsar, \src, possibly associated with the supernova remnant \srce in the Large Magellanic Cloud, using \xmm X-ray observations. Pulsations with a periodicity of 570\,s are discovered from the Be X-ray binary \src confirming its nature as a HMXB pulsar. The HMXB is located near the geometric centre of the supernova remnant \srce (0.9\arcmin\ from the centre) which supports the XRB-SNR association. The estimated age of the supernova remnant is 43--63\,kyr years which points to a middle aged to old supernova remnant. The large diameter of the supernova remnant combined with the lack of distinctive shell counterparts in optical and radio indicates that the SNR is expanding into the tenous environment of the superbubble N103. The estimated magnetic field strength of the neutron star is $B\gtrsim10^{14}$\,G assuming a spin equilibrium condition which is expected from the estimated age of the parent remnant and assuming that the measured mass-accretion rate remained constant throughout.

\end{abstract}




\section{Introduction}
A neutron star (NS) X-ray binary associated with its parent supernova remnant (SNR) is an extremely rare object and can provide unique insights on the early evolutionary stages of NSs in the presence of a binary companion. The visibility time of a  SNR is only a few $10^4$~yr, which is typically three orders of magnitude shorter than the lifetime of high-mass X-ray binaries (HMXBs). 
A HMXB-parent SNR association therefore implies a very young binary system. A majority of these associations have been found in the Magellanic Clouds (MCs) in recent years, given their ideal environment for hosting young stellar remnants, a high formation efficiency for high-mass X-ray binaries (HMXBs), as well as relatively small distance and low foreground absorption conducive to performing detailed studies. 

Discovered XRB-SNR associations include LXP\,4.4 \citep{2019MNRAS.490.5494M}, SXP\,1062 \citep{2012A&A...537L...1H,2018MNRAS.475.2809G}, CXO J053600.0$-$673507 \citep{2012ApJ...759..123S}, SXP\,1323 
\citep{2019MNRAS.485L...6G} in the Magellanic Clouds, and SS\,433 and Circinus\,X-1 in our Galaxy \citep{1980A&A....84..237G,2013ApJ...779..171H}. The youngest among them until now are Circinus\,X-1, with an estimated age $<$4600\,years \citep{2013ApJ...779..171H} and LXP\,4.4 with an estimated age of $<$6000\, years \citep{2019MNRAS.490.5494M}. 

\srce is a  candidate SNR in the LMC \citep{2017ApJS..230....2B} that was first reported by \citet{2000AJ....119.2242C} as RX\,J050736$-$6847.8, a large ring ($\sim$ 150 pc) of diffuse X-ray emission projected in the vicinity of the superbubble LHA 120-N~103 (hereafter N103). Superbubbles are large structures in the interstellar medium created by the supernova explosions of massive stars and their stellar winds in an OB association or stellar cluster. The shock heated gas in superbubbles emit in X-ray wavelengths. Detection of excess of diffuse X-ray emission in superbubbles are indicative of the presence of interior SNRs shocking the inner walls of the superbubble shell \citep[see for e.g.][]{2001ApJS..136..119}. 

The above category of SNRs expand in the low-density medium of the superbubble, and have very weak optical and radio emission associated with them. Therefore, the nature of these systems cannot be confirmed using the conventional SNR diagnostics i.e. presence of a high  \SII\ / \Halpha\ line ratio and non-thermal radio emission coincident in X-ray emission.
The candidate \srce was likewise indicated to be the largest SNR in the LMC expanding in the low density environment of the superbubble N103, and hence with no discernible optical emission (Magellanic Clouds Emission Line Survey) and radio continuum emission \citep{2000AJ....119.2242C,2017ApJS..230....2B,2020MNRAS.500.2336Y}. The X-ray luminosity of the system lies within the range expected for SNRs, and the age was estimated to be $\sim5\times10^{4}$ yr based on the Sedov solution \citep{2000AJ....119.2242C}. 

In this work we identify for the first time the BeXRB \src and \srce as a possible SNR-HMXB association and investigate the properties of the SNR and its compact object in detail. 
We report the discovery of pulsations from the BeXRB \src located near the geometrical centre of the SNR candidate \srce. 
This confirms its nature as an NS.  \citet{2018MNRAS.475.3253V} identified the source as a BeXRB in the LMC with the optical companion classified as a B3 IIIe star and suggested a binary orbital period of 5.27\,days. 
We identified a more likely orbital period of 40.2\,days using more than 22 years of OGLE monitoring data.
The observations and their analysis are described in Section~\ref{Sect:data}. Section~\ref{Sect:results} presents the results and Section~\ref{Sect:discussion} the discussion and Section~\ref{Sect:conclusion} the conclusions.

\section{Observations and analysis}
\label{Sect:data}

\begin{table*} 
\caption{\xmm observations details of \src.} 
\begin{tabular}{ccccc} 
\hline
Date       & ObsID      & Exposure           & Off-axis angle         & Telecope vignetting\\ 
           &            & PN / MOS2 / MOS1   & PN / MOS2  / MOS1      & PN / MOS2 / MOS1 \\
yyyy/mm/dd &            & (ks)               &                        &                   \\ 
\hline 
2000/07/07 & 0113000301 & 11.8 / 16.2 / 23.6 &  10.4\arcmin / 9.7\arcmin / 8.7\arcmin & 0.51 / 0.48 /  0.52 \\ 
2017/10/19 & 0803460101 & 53.0 / 55.0 / 55.0 &  10.0\arcmin / 11.1\arcmin / 11.1\arcmin & 0.52 / 0.54 /  0.55 \\ 
\hline 
\label{tabxray} 
\end{tabular} 
\end{table*} 

\subsection{X-ray observations and analysis}
\label{xray}
\src was observed serendipitously with \xmm  twice in 2000 (Obs1 from now) and again in 2017 (Obs2 from now). The observation details are given in Table~\ref{tabxray}. EPIC \citep{2001A&A...365L..18S,2001A&A...365L..27T} observations were processed with the \xmm data analysis software SAS version 18.0.0\footnote{Science Analysis Software (SAS): http://xmm.esac.esa.int/sas/}. We searched for periods of high background flaring activity by extracting light curves in the energy range of 7.0--15.0\,keV and removed the time intervals with background rates $\geq$~8 and 2.5\,cts\,ks$^{-1}$~arcmin$^{-2}$ for EPIC-PN and EPIC-MOS respectively \citep{2013A&A...558A...3S}. Events were extracted using the SAS task \texttt{evselect} by applying the standard filtering criteria (\texttt{\#XMMEA\_EP \&\& PATTERN<=4} for EPIC-PN and \texttt{\#XMMEA\_EM \&\& PATTERN<=12} for EPIC-MOS).

The \swift/XRT data were analysed following the \swift data analysis guide\footnote{\url{http://www.swift.ac.uk/analysis/xrt/}} \citep{2007A&A...469..379E}. Because of the low flux of \src in all observations, we only performed a simple source detection and position determination using a sliding-cell detection algorithm implemented by {\tt XIMAGE and sosta}\footnote{\url{https://heasarc.gsfc.nasa.gov/xanadu/ximage/ximage.html}}. 
For non-detections we estimated the 3$\sigma$ upper limits using a Bayesian method introduced by \citet{1991ApJ...374..344K}.

\begin{figure*}
 \includegraphics[scale=0.38]{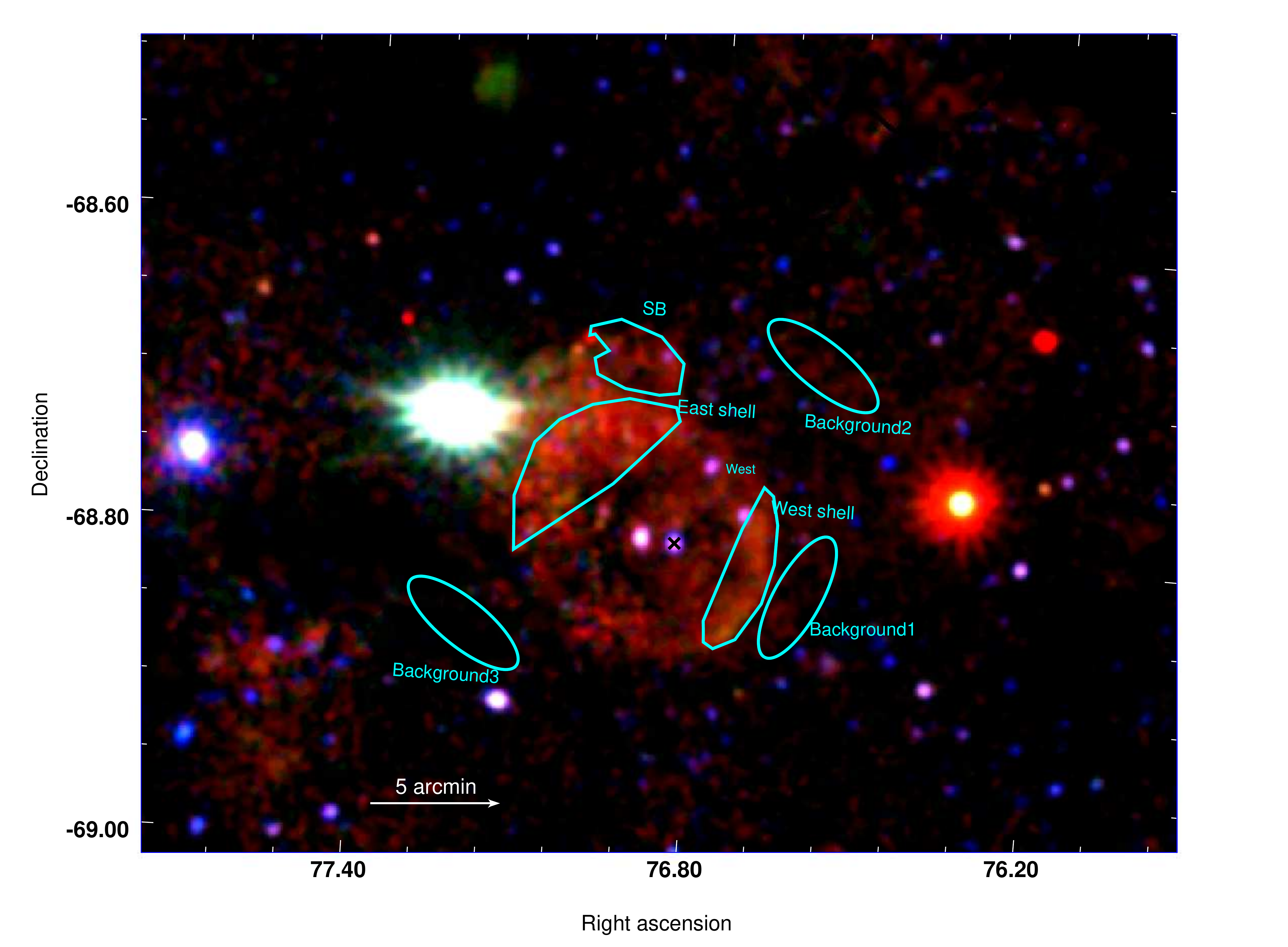}
   \caption{\xmm EPIC RGB (R=~0.3--0.7\,keV, G=~0.7--1.1\,keV, B=~1.1--4.2\,keV) image of \src. The regions used for spectral extraction are shown in cyan.
   The black cross shows the optical position of the BeXRB.} 
   \label{fig1}
\end{figure*}

\begin{figure*}
 \includegraphics[scale=0.45]{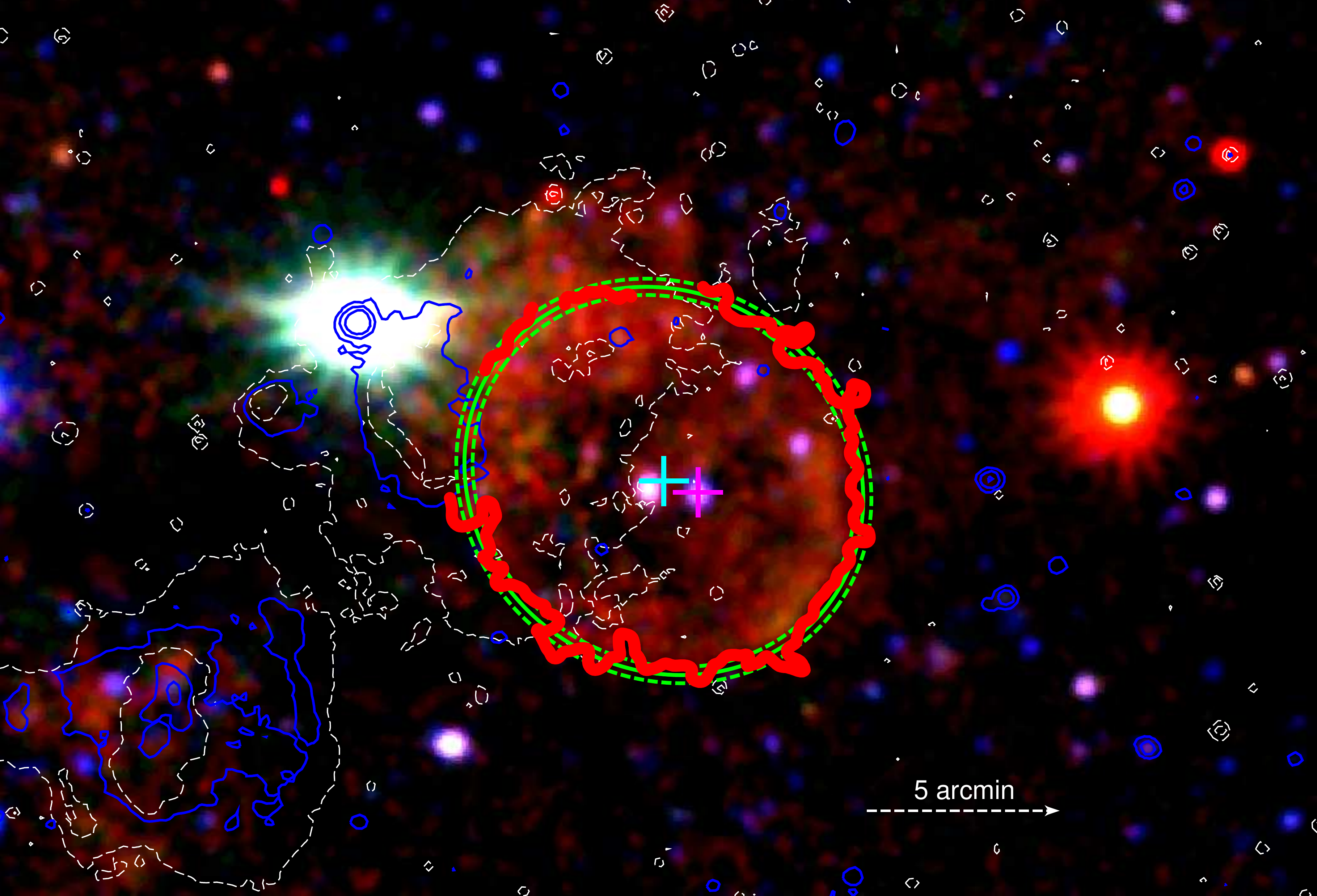}
   \caption{Zoomed in \xmm EPIC RGB (R=~0.3--0.7\,keV, G=~0.7--1.1\,keV, B=~1.1--4.2\,keV) image of \src. Overlaid in blue are radio contours from the latest Australian Square Kilometre Array Pathfinder (ASKAP) survey of the LMC at 888~MHz (bandwidth is 288~MHz). The radio continuum contours correspond to 1, 2, and 3~mJy~beam$^{-1}$ while the image beam size is 13.7\arcsec$\times$11.8\arcsec\ and local rms is $\sim$0.2~mJy~beam$^{-1}$. 
   The white line overlays the H$\alpha$ image contours from the Magellanic Clouds Emission Line Survey (MCELS ) \citep{2004AAS...20510108S}.
   The cyan cross shows the best-fit centre of the SNR and the magenta cross the position of the optical counterpart of the BeXRB. 
 The red solid line indicate the X-ray contour level corresponding to 3$\sigma$ above the average background
surface brightness. The green solid line shows the best-fit ellipse to the contour, with the dashed lines denoting the 1$\sigma$ errors on the
best-fit.} 
   
 \label{fig2}
\end{figure*}


\begin{table} 
\caption{Swift observations details of \src.} 
\begin{tabular}{ccccc} 
\hline
Date  & ObsID  & Mode  & Exposure & Observed \\ 
      &        &       &  (s)     & Luminosity \\ 

55962.8 & 00045577002 & PC & 2128 & <8.6 \\
55994 & 00045444002 & PC & 248 & <77.3 \\
56252.5 & 00045444003 & PC & 779 & <37.8 \\
56254.9 & 00045444004 & PC & 904 & <28.8 \\
56259.3 & 00045444005 & PC & 1116 & 56.7 \\
56377.8 & 00045576001 & PC & 2872 & <6.9 \\
56968.9 & 00033486001 & PC & 961 & <14.4 \\
56972.2 & 00033486002 & PC & 434 & <31.8 \\
57109.2 & 00033486003 & PC & 439 & <31.5 \\
57653 & 00034730001 & PC & 1101 & <18.4 \\
58220.5 & 00094089001 & PC & 247 & <55.9 \\
58234.3 & 00094089003 & PC & 243 & <57.1 \\
58255.6 & 00094089004 & PC & 245 & <56.5 \\
58262.3 & 00094089005 & PC & 247 & <55.9 \\
58290.9 & 00094089007 & PC & 245 & <56.5 \\
58332.6 & 00094089009 & PC & 250 & <55.4 \\
58346.8 & 00094089010 & PC & 245 & <56.5 \\
58360.7 & 00094089011 & PC & 232 & <59.5 \\
58374.7 & 00094089012 & PC & 250 & <55.4 \\
58388.5 & 00094089013 & PC & 240 & <57.7 \\
58416.8 & 00094089014 & PC & 242 & <57.1 \\
59006.9 & 00034730003 & PC & 1925 & <11.6 \\
\hline 
\label{tabxrayswift} 
\end{tabular}

\tnote{a)} \footnotesize{Luminosity in the 0.5-10\,keV band in units of  10$^{34}$ erg s$^{-1}$.}
\tnote{b)} \footnotesize{To convert \swift/XRT count rates to luminosities we used the spectral parameters obtained from the fit to the \xmm data. The conversion factor is 2.1$\times 10^{37}$ erg s$^{-1}$ (c/s)$^{-1}$. Only \swift exposures of $\geq$200\,s covering the source have been used for this purpose.}
\end{table} 

\subsection{OGLE}
\label{ogle}

The infrared counterpart of the HMXB \src is identified to be 
\osrc with $J$, $H$ and $K$ magnitudes of 15.0, 15.3 and 14.6 mag respectively. \osrc was observed by the Optical Gravitational Lensing Experiment (OGLE), which started observations in 1992 \citep{1992AcA....42..253U} and continued observing until today \citep[OGLE-IV,][]{2015AcA....65....1U}, but interrupted since March 2020. {\bf Optical observations} were performed with the 1.3\,m Warsaw telescope at Las Campanas Observatory, Chile. Images are taken in the V and I filter pass-bands and photometric magnitudes are calibrated to the standard VI system.

\section{Results}
\label{Sect:results}

\subsection{Morphology of \srce}
Figures~\ref{fig1} and \ref{fig2} display the combined \xmm EPIC  image centered on \srce overlaid with optical (MCELS) and radio (ASKAP) contours. The source morphology resembles a large circular shell-like structure of diffuse emission of $\sim$150 pc in diameter. It is projected in the vicinity of the superbubble N103 adjacent to SNR N103B (MCSNR J0508-6843) in the east \citep{2016A&A...585A.162M,2017ApJS..230....2B}. The north-eastern part of the shell overlaps with the hot gas of the superbubble which is also indicated by the presence of strong optical H$\alpha$ and radio-continuum emission coincident with this region. The south-western shell is more clearly defined.
No optical or radio emission is detected from the shell region which is possibly due to the fact that \srce expands in the low-density environment of the superbubble.
To measure the size of the shell, we employed the method described by \citet{2015A&A...579A..63K}, which fits an ellipse to the outer contours of the shell (at 3$\sigma$ above the surrounding background level in the 0.2$-$1 keV EPIC image). Due to the contamination of the northern shell region with the SB N103 as seen from the H$\alpha$ contours in Fig.~\ref{fig2}, the extended emission from the north is used as a background component to define the contour around the northern shell. We derive the ellipse centre at R.A. = 05$^{\rm h}$07$^{\rm m}$32.1$^{\rm s}$ and Dec. = $-$68\degr47\arcmin40.7\arcsec (J2000). The semi-major and semi-minor axes of $5.32\arcmin (\pm0.21 \arcmin)$ and $4.84\arcmin (\pm0.21 \arcmin)$, respectively, correspond to a size of 154.6 $\times$ 140.8 ($\pm6.1$)~pc at the distance of the LMC (50~kpc) with the major axis rotated $\sim$49.3\degr\ East of North. 


\subsection{Identification of point sources inside the shell}
In order to identify the point sources, we performed a maximum-likelihood source detection analysis on the \xmm/EPIC images on fifteen images created from the three EPIC cameras in five energy bands as given in \citep{2009A&A...493..339W,2013A&A...558A...3S}. Source detection was performed simultaneously on all the images using the SAS task {\tt edetect\_chain}.
Two point-like sources are found near the centre of the shell morphology.
The source to the {\bf east} is identified as a spectroscopically-confirmed quasar MQS J050736.44$-$684751.6 with $z=0.53$ \citep{2012A&A...537A..99S,2013ApJ...775...92K} and is hence a background object which is projected in the line of sight of \srce.

The other point source marked with black cross in Fig.~\ref{fig1} and magenta cross in Fig.~\ref{fig2} lies 0.9\arcmin\ away from the geometric centre of the SNR. Its best-determined position is R.A. = 05$^{\rm h}$07$^{\rm m}$22.37$^{\rm s}$ and Dec. = $-$68\degr47\arcmin58.2\arcsec (J2000) with a $1\sigma$ statistical uncertainty of 0.72\arcsec. The positional error is dominated by systematic astrometric uncertainties and a systematic error of 0.37\arcsec\ was added in quadrature \citep{2016A&A...590A...1R}.  The source was already identified as a  BeXRB in the LMC from optical spectroscopic observations \citep[BeCand 3 in][]{2018MNRAS.475.3253V}. The confirmation of the nature of the source as a neutron star (see later section) and its positional proximity to the geometric centre of the SNR indicates that it is the compact object born out of the explosion of \srce. We calculated the probability of chance coincidence for a HMXB pulsar to lie within 0.9\arcmin\ of the centre of an SNR within the LMC. For this we used the total of 21 known HMXBs with detected pulsations and $\sim$57 confirmed SNRs (with sizes that can be resolved with \xmm) to be identified within the \xmm observations of the LMC which cover an area of $\sim$20 deg$^{2}$ in total. Assuming that the HMXBs and SNRs are uniformly distributed within the survey area, the probability of finding a HMXB by chance within 0.9\arcmin\ of the centre of an SNR is 0.04. It is to be noted that HMXBs and core-collapse SNRs \citep[which constitute about 60\% of all LMC SNRs,][]{2016A&A...585A.162M} are not uniformly distributed but follow star-forming regions, so this probability is likely underestimated. The probability of chance coincidence would also be slightly higher if all known HMXBs in the LMC are taken into account. The low probability supports the association of the HMXB with the SNR, but a chance coincidence can formally not be ruled out.

\subsection{OGLE monitoring of the optical counterpart}

\subsubsection{Long-term OGLE I-band light curve}

\begin{figure*}
   \centering
   \resizebox{0.9\hsize}{!}{\includegraphics[clip=]{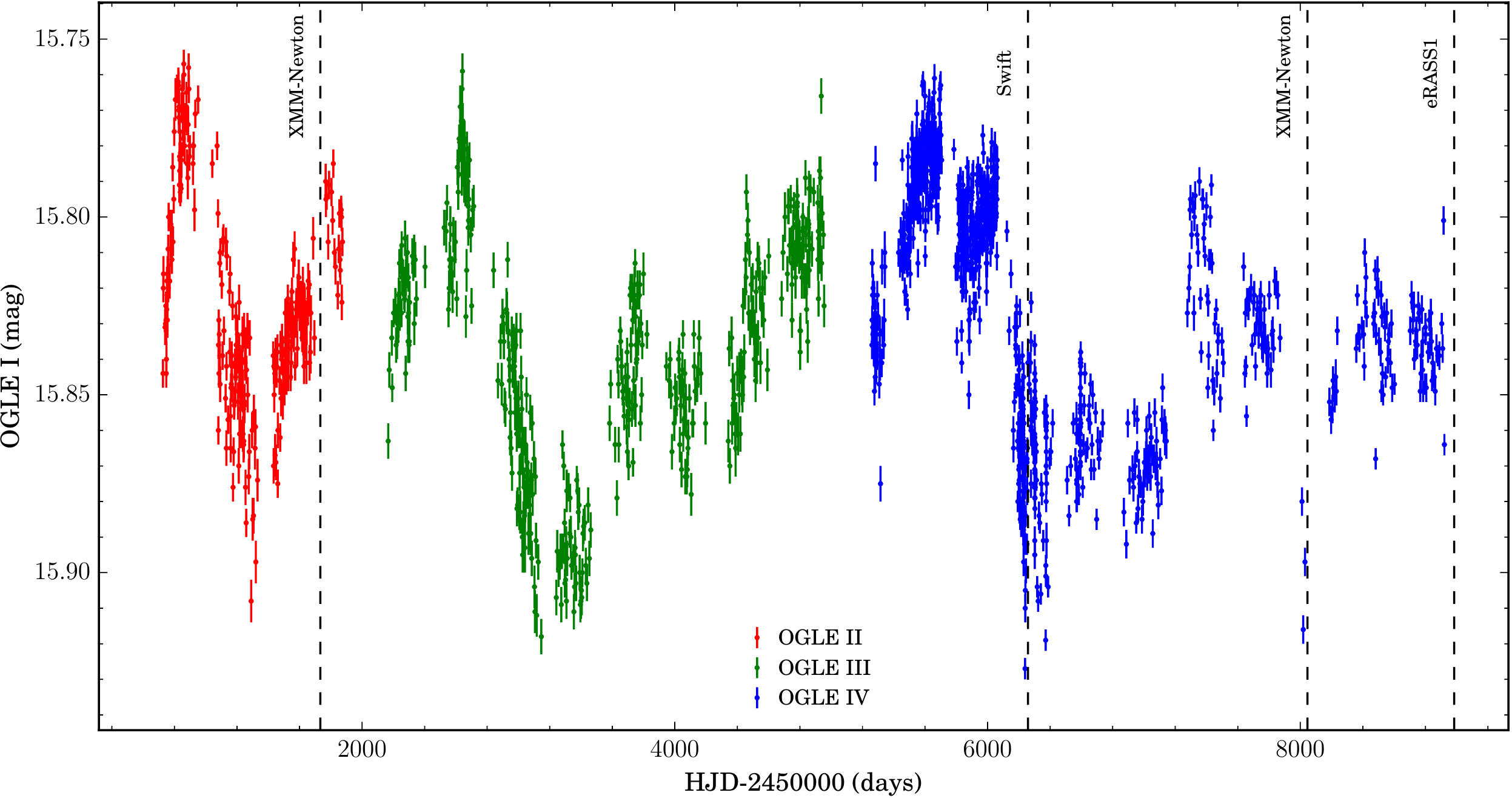}}
   \caption{OGLE I-band light curve of \osrc from October 1997 to March 2020. Epochs of X-ray observations analysed in this work are marked with vertical dashed lines.}
   \label{ogle-lc-iband}
\end{figure*}

\begin{figure}
   \centering
   \resizebox{0.95\hsize}{!}{\includegraphics[clip=]{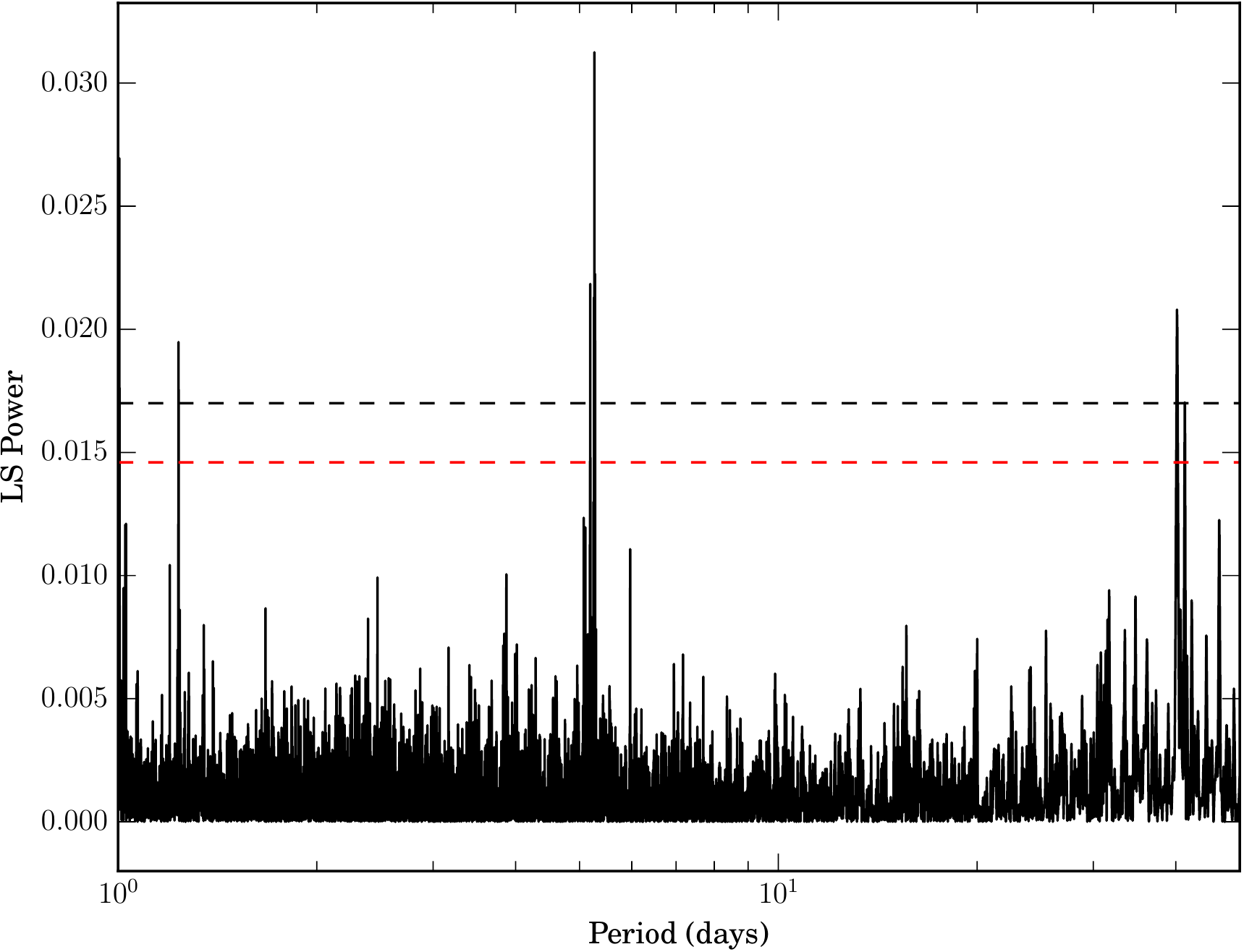}}
   \caption{Lomb-Scargle periodogram of the detrended OGLE I-band light curve of \osrc. Significant peaks are found at 1.235, 5.266 and 40.16\,days. The red and black dashed lines mark the 95\% and 99\% confidence levels, respectively.}
   \label{ogle-lc-ls}
\end{figure}

\begin{figure}
   \centering
   \resizebox{0.95\hsize}{!}{\includegraphics[clip=]{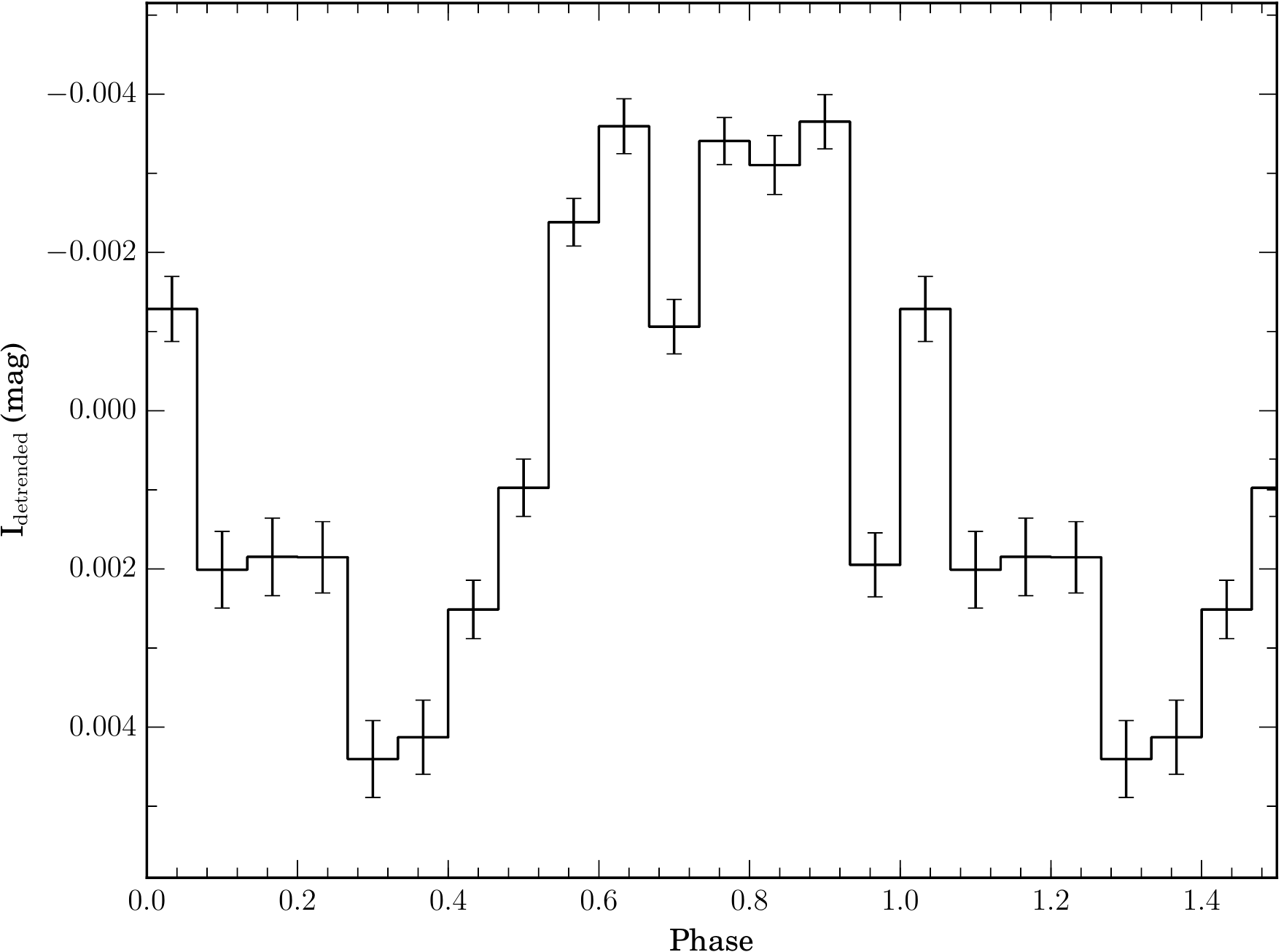}}
   \caption{Detrended OGLE I-band light curve folded with a period of 40.1609\,days.}
   \label{ogle-lc-fold}
\end{figure}

Figure~\ref{ogle-lc-iband} shows the OGLE I-band light curve of the optical counterpart of \src obtained during observing phases II-IV over a period of 13 years. The light-curve is variable in nature showing dip-like features, typical of Be stars with a binary companion. The average I band magnitude of the source is 15.8 mag. 
In order to verify the orbital period of the system, we detrended the light curve by subtracting a smoothed light curve \citep[derived by applying a Savitzky-Golay filter with a window length of 101 data points,][]{1964AnaCh..36.1627S} and computed Lomb-Scargle peridograms \citep{1976Ap&SS..39..447L,1982ApJ...263..835S}. A highly significant peak is found in the periodogram at 5.266\,days with two other peaks at 1.235 and 40.16\,days (Fig.~\ref{ogle-lc-ls}). The 1.235 and 5.266\,day periods (frequencies of 0.810 and 0.190\,day$^{-1}$, respectively) are aliases of each other with the 1\,day sampling period of the light curve. Since short periods near one day are most likely caused by non-radial pulsations of the Be star \citep[e.g. ][]{2013MNRAS.431..252S}, we suggest the 40.16\,day period as the most likely orbital period of the system, which is also more typical for BeXRBs \citep{2016A&A...586A..81H}. 
The detrended light curve folded with that period is presented in Fig.~\ref{ogle-lc-fold}.

\subsubsection{OGLE V-band and colour variations}

\begin{figure}
   \centering
   \resizebox{0.95\hsize}{!}{\includegraphics[clip=]{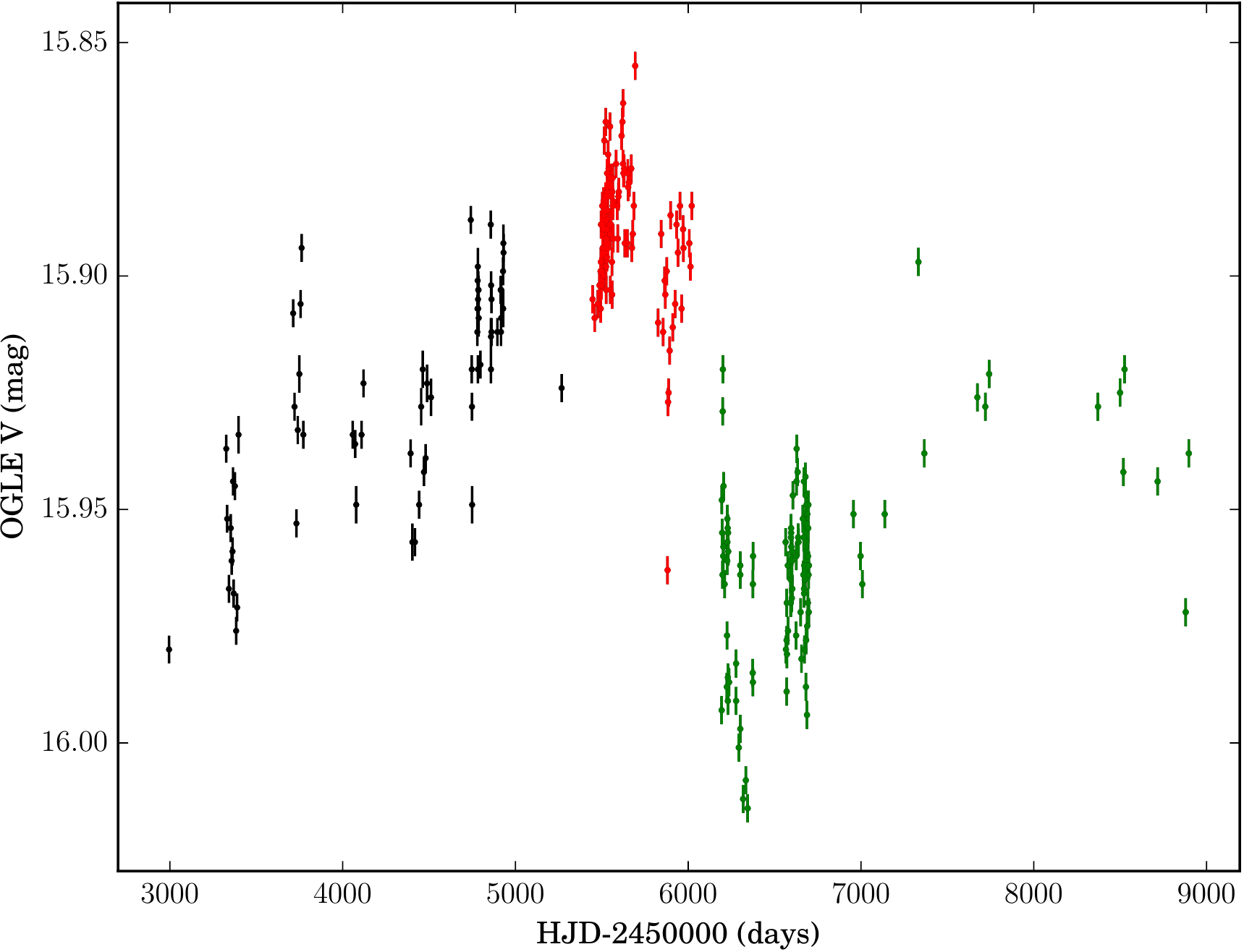}}
   \caption{OGLE V-band light curve of \osrc. The colours mark different phases of brightness and colour evolution as shown in Fig.\,\ref{ogle-lc-color}.}
   \label{ogle-lc-vband}
\end{figure}

\begin{figure}
   \centering
   \resizebox{0.95\hsize}{!}{\includegraphics[clip=]{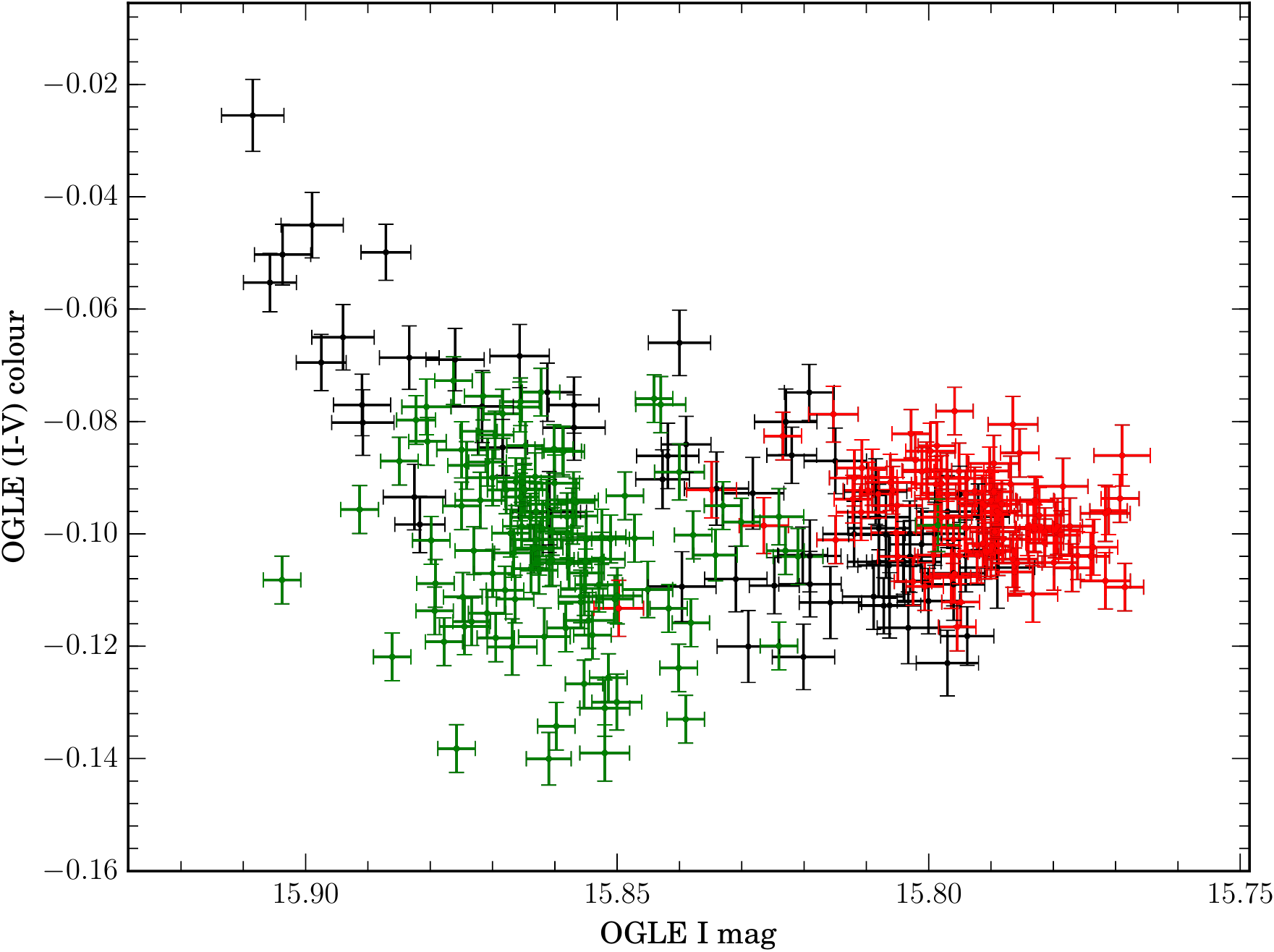}}
   \caption{OGLE colour (I-V) - magnitude (I) diagram of \osrc. During brightness rise (black) the emission becomes redder reaching asymptotically a minimum value in I-V (red), which is also maintained after the brightness drop (green).}
   \label{ogle-lc-color}
\end{figure}

During OGLE phases III and IV also V-band measurements of the optical counterpart of \src are available, in particular around 
the sharp drop in brightness around MJD 56200\,d (Figs.\,\ref{ogle-lc-iband} and \ref{ogle-lc-vband}).
We created I-V colour indices by using the measured V-band magnitudes and neighbouring I-band values interpolated to the time of the V-band measurement. The colour - magnitude diagram is shown in Fig.\,\ref{ogle-lc-color}.

\subsection{X-ray timing analysis}
\begin{figure}
   \centering
   \includegraphics[width=1.0\columnwidth]{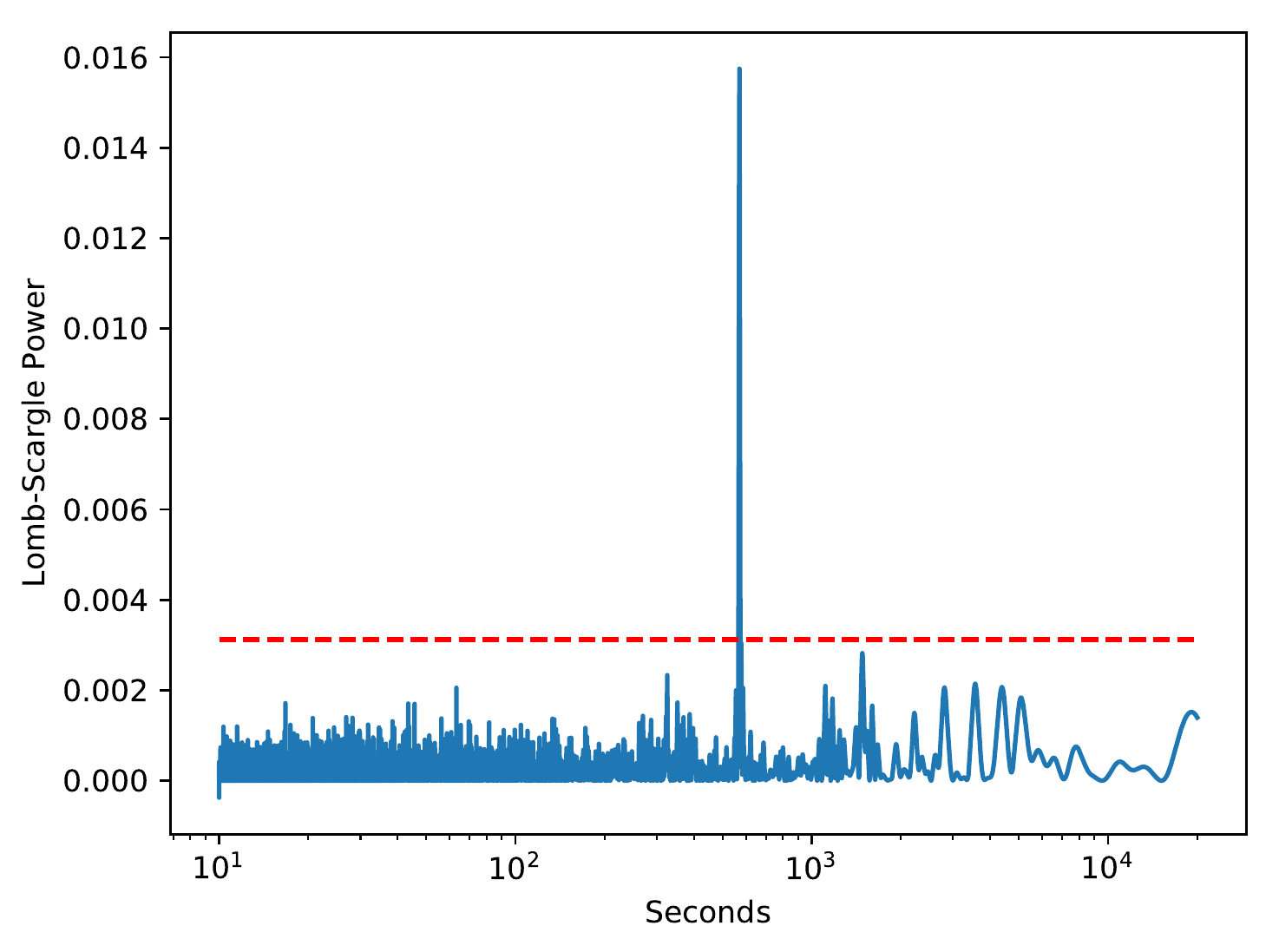}
   \caption{Lomb-Scargle periodogram of the \xmm EPIC light curve in the energy band of 1--12\,keV (ObsID 0803460101). The peak indicates the spin period of the neutron star. The red  dashed line mark the 99.73\% confidence levels.
   }
   \label{figtiming}
\end{figure}

\begin{figure}
\centering
\includegraphics[width=0.7\columnwidth,angle=-90]{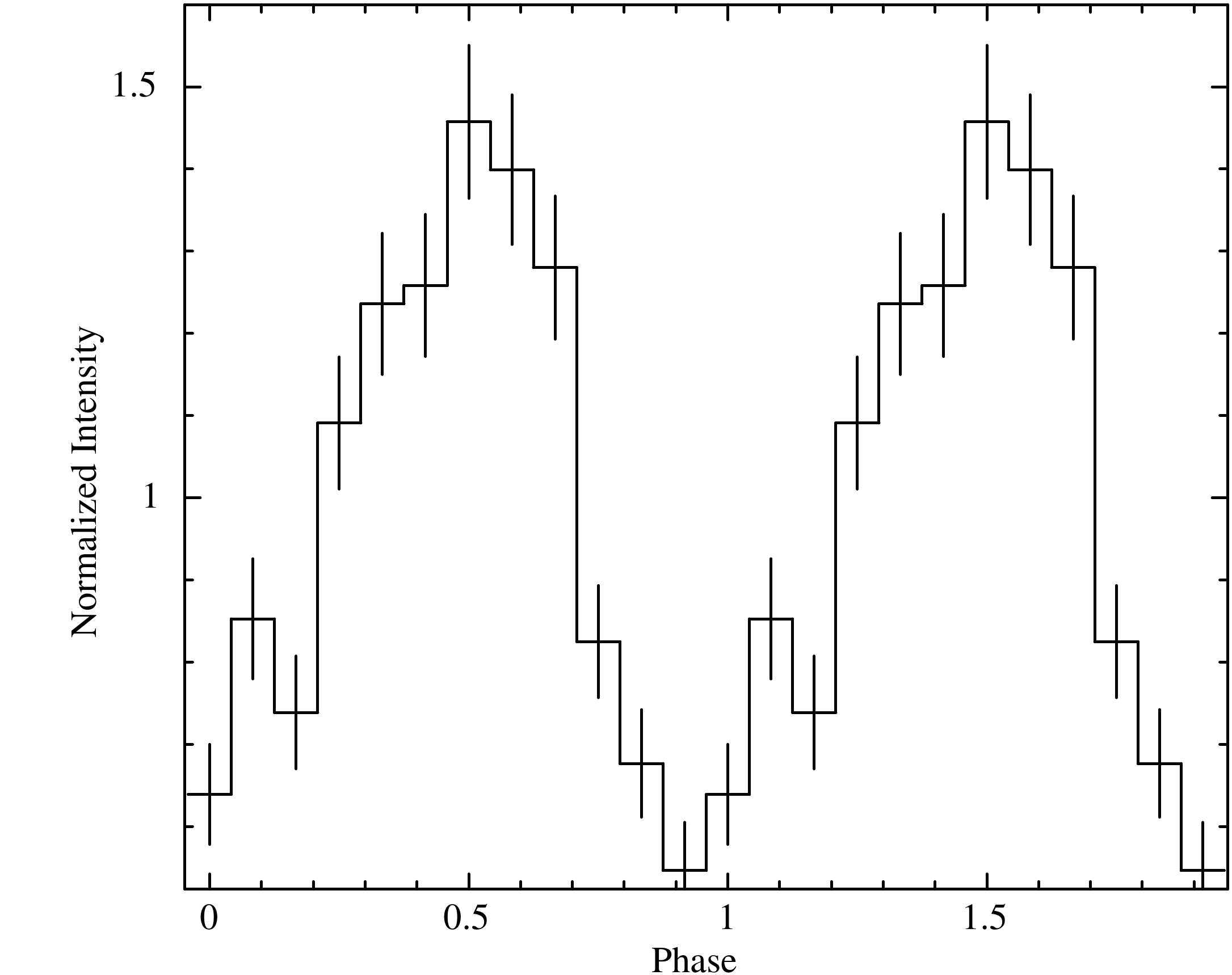}
\caption{Corrected EPIC-PN light curve folded with 570.35\,s, showing the pulse profile of the HMXB in the energy band of 1--12\,keV.}
\label{figpp}
\end{figure}

\src displayed a net count rate (PN) of 0.03 c/s during Obs. 1 and 0.05 c/s during Obs. 2 in the energy range of 0.2--12\,keV confirming its variable nature. The X-ray light curves during the individual observations however did not exhibit variability on shorter time scales. Since the source was brighter during Obs. 2 and was observed for a longer duration, data from this observation was considered to study the temporal properties of the source in detail. To look for a possible periodic signal in the X-ray light curve of the HMXB, we extracted source events using a circular region with radius 26\arcsec\ centred on the best-fit position, and a background region of a larger size away from the source as shown in Fig.~\ref{fig1}. The light curve was corrected for all effects like vignetting and Point Spread Function losses by the task \texttt{epiclccorr}.
At first, we searched for a periodic signal in the barycentre-corrected \xmm EPIC light curve in the energy range above $1$\,keV using a Lomb-Scargle periodogram analysis in the period range of 0.5-3000\,s \citep{1976Ap&SS..39..447L,1982ApJ...263..835S}. A strong periodic signal is detected at 570.4\,s indicating the spin period of the neutron star in the BeXRB (Fig.~\ref{figtiming}). 
In order to determine the pulse period more precisely, we employed the Bayesian periodic signal detection method described by \citet{1996ApJ...473.1059G}.
 The spin period and its associated 1$\sigma$ error are determined to  570.35$\pm$0.35\,s.
The spin period and its associated \xmm EPIC light curve in the range of 1--12\,keV, folded with the best-obtained period is shown in Fig.~\ref{figpp}.
The pulse fraction in the same energy range is 40\% and no change in the pulse shape or pulse fraction can be detected within this energy range. Pulsations cannot be detected below 1~keV, possibly due to the contamination from the SNR.

The source was scanned 49 times during the first all-sky survey (eRASS1) of the eROSITA instrument \citep[][]{Predehl2020} on board the Russian/German Spektrum-Roentgen-Gamma (SRG) mission. The scans spanned between MJD 58980.53 to MJD 58987.09, accumulating a total exposure of 1543 s. The source was variable during the scans and the count rates {\bf remained around} 0.08$\pm$0.03 c/s (0.2-8.0\,keV) around MJD 58981.54 in the beginning and dropped to $<$ 0.01 c/s towards the end of the eRASS1 scans.
Source detection was performed simultaneously on all the images in the standard eROSITA energy bands of 0.2--0.6\,keV, 0.6--2.3\,keV and 2.3--5.0\,keV.
The vignetting and Point Spread Function corrected count rate in the energy range of 0.2-5.0\,keV was  0.05$\pm$0.01 c/s indicating a similar average flux as observed during \xmm Obs. 2.

\subsection{X-ray spectral analysis}
For the spectral analysis, the SAS tasks \texttt{rmfgen} and \texttt{arfgen} were used to create the redistribution matrices and ancillary files. The significant extent of the SNR and varying off-axis position in the two observations was taken into account by extracting spectra from vignetting-weighted event lists, created through the SAS task \texttt{evigweight} \citep[as described in][]{2016A&A...585A.162M}. To account for the spatially-dependent non-X-ray background (NXB) in extended emission spectra, spectra were extracted from Filter Wheel Closed (FWC) data at the same detector position.
Spectra were binned to achieve a minimum of 20 and 25 counts per spectral bin for the point source and SNR, respectively, to allow using the $\chi^{2}$-statistic. The spectral analysis was performed using the {\small XSPEC} fitting package, version~12.9 \citep{1996ASPC..101...17A}. Errors were estimated at the 90\,percent confidence level, unless otherwise stated.

\begin{table*}
\begin{threeparttable}[t]

    \caption{X-ray spectral parameters of \srce}
    \label{tabsnrspec}

    \begin{tabular}{l  @{\hspace{1.0em}}  ccccccccc}
    \hline
    Region & $N_H ^{{\rm LMC}}$ & $kT$ & $\tau = n_e \,t$ & O & Ne & Mg & Fe & EM\tnote{a} & $\chi^2/$dof ($\chi^2 _r$)\\
    \noalign{\smallskip}
           & (10$^{20}$ cm$^{-2}$) & (keV) & ($10^{11}$~cm$^{-3}$~s$^{-1}$) &  &  &  &  & ($10^{57}$~cm$^{-3}$)\\
    \hline
     & \multicolumn{8}{c}{\texttt{vapec} parameters} \\
   \hline
    Shell & 7.1$_{-3.5} ^{+4.2}$ & 0.25$\pm0.01$ & --- & 0.27$_{-0.07} ^{+0.10}$ & 0.38$_{-0.10} ^{+0.16}$ & 0.63$_{-0.24} ^{+0.37}$ & 0.12$_{-0.03} ^{+0.05}$ & 1.21$_{-0.38} ^{+0.48}$ & 2156.4/2051 (1.05) \\
    \noalign{\smallskip}
    East  & 0.8 $(< 5.5)$ & 0.22$\pm0.01$ & --- & \multicolumn{4}{c}{---} & 0.98$_{-0.11} ^{+0.29}$ & 154.7/140 (1.11)\\
    \noalign{\smallskip}
    West  & 3.3 ($< 12.7$)& 0.23$\pm0.01$ & --- & \multicolumn{4}{c}{---} & 0.89$_{-0.18} ^{+0.40}$ & 225.3/213 (1.06)\\
    \noalign{\smallskip}
    SB    & 0.0 ($<7.6$)  & 0.23$\pm0.03$ & --- & \multicolumn{4}{c}{---} & 0.53$\pm0.12$ & 92.5/88 (1.05)\\
    \hline
    & \multicolumn{8}{c}{\texttt{vpshock} parameters\tnote{b}}\\
   \hline
    Shell & 0.0 ($< 1.3$) & 0.36$_{-0.03} ^{+0.07}$ & 6.3$_{-3.4} ^{+8.1}$ & 0.30$_{-0.05} ^{+0.04}$ & 0.40$_{-0.03} ^{+0.07}$ & 0.44$\pm0.16$ & 0.11$\pm0.02$ & 0.49$_{-0.12} ^{+0.13}$ & 2148.6/2050 (1.05) \\
    \noalign{\smallskip}
    East  & 0.0 ($<1.6$) & 0.77$_{-0.21} ^{+0.43}$ & 0.3$_{-0.1} ^{+0.2}$ &  \multicolumn{4}{c}{---} & 0.14$_{-0.02} ^{+0.03}$& 149.9/139 (1.08)\\
    \noalign{\smallskip}
    SB    & 0 ($<39.9$) & 0.88$_{-0.46} ^{+1.57}$ & 0.2$_{-0.1} ^{+0.6}$ &  \multicolumn{4}{c}{---} & 0.07$_{-0.02} ^{+0.51}$ & 90.2/87 (1.04)\\
    \hline
    \end{tabular}

  \begin{tablenotes}
  \item[a] The emission measure $EM = n_e n_{{\rm H}} V$, the product of electronic and proton densities with the total emitting volume, acts as a (temperature-dependent) normalisation of thermal plasma models.
  \item[b] The ionisation time scale could not be constrained for the West region, and thus we do not report \texttt{vpshock} parameters for it.
  \end{tablenotes}

\end{threeparttable}
\end{table*}

\subsubsection{\srce}
\label{analysis_spectral_SNR}
Spectra were extracted from the entire shell region, from the eastern and western hemispheres, and from the superbubble region to the North (see Fig.~\ref{fig1}). Background spectra were accumulated in nearby regions chosen individually per instrument and observation. In Obs.~1 we also made sure to exclude a large region around the bright SNR N103B and its streak of out-of-time events. Point sources detected in the background and shell region, including the background AGN and HMXB, were excised using circular regions with a 25\arcsec radius. Thus, we estimated that $<5\%$ of the source counts from these sources will {\bf contaminate} our extended source spectra due to PSF leaks. Source and background spectra were fitted simultaneously, with NXB parameters tied to those obtained from the FWC spectra, and astrophysical X-ray background (AXB) parameters constrained from background spectra. The AXB comprises Galactic thermal emission (Local Hot Bubble, Galactic halo), cosmic X-ray background, and LMC diffuse emission. More details on this method can be found in \citet{2016A&A...585A.162M,2019A&A...631A.127M}.

The X-ray absorption was modeled using the {\texttt tbabs} model \citep{2000ApJ...542..914W} with atomic cross sections adopted from \citet{1996ApJ...465..487V}. We used two absorption components: The first one to describe the Galactic foreground absorption, where we used a fixed column density of $6\times 10^{20}$~cm$^{-2}$ \citep{1990ARA&A..28..215D} with abundances taken from \citet{2000ApJ...542..914W}. The second component accounts for the LMC material in front of the object. For the latter absorption component, the abundances were set to LMC abundances following \citet{2016A&A...585A.162M} and the column density $N_H ^{{\rm LMC}}$ was free in the analysis. The cosmic X-ray background, {\bf seen} through all the LMC, was absorbed by the total line-of-sight column density of $2.0\times 10^{21}$~cm$^{-2}$, obtained from the ATCA+Parkes map of \citet{2003ApJS..148..473K} averaged over 0.1 squared degree.


\begin{figure}
\includegraphics[angle=-90,width=\hsize]{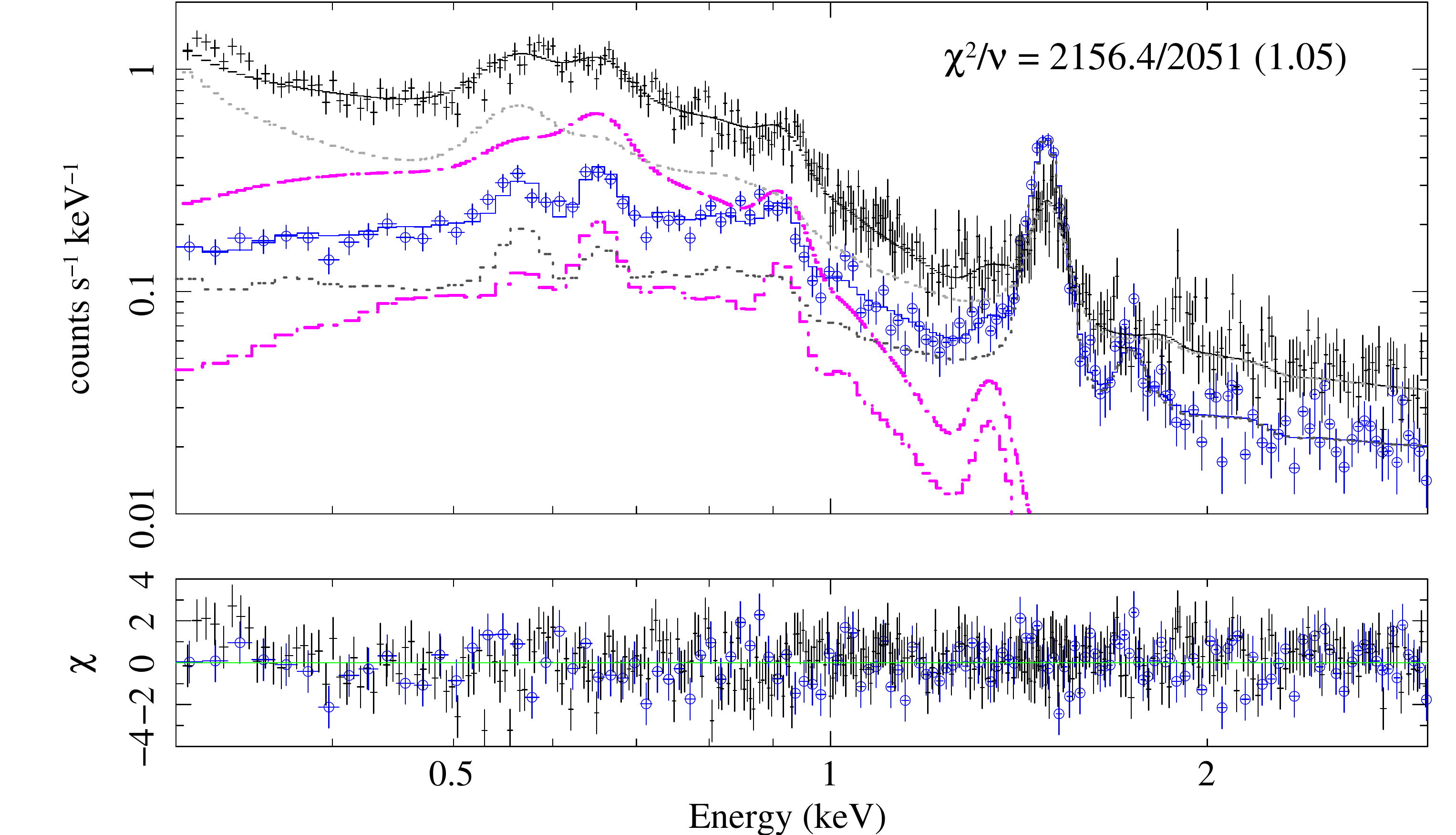}
\caption{X-ray  spectra  of  the  SNR  \srce. 
The upper panel shows the simultaneous spectral fit using all the available data (Table~\ref{tabxray}). For the sake of clarity only  PN and MOS2 data from obsid 0803460101 are displayed in black dots and blue circles, respectively.
The total background models for PN and MOS are shown by the dotted  gray  lines.  The best-fit \texttt{vapec} model for the SNR emission is  shown  by  the dashed magenta lines, as convolved with the PN and MOS spectral responses (top and bottom curves, respectively).
The  lower  panel  displays  the  residuals after the fit.
}
\label{figspecsnr}
\end{figure}

We fit the X-ray spectra with collisional ionisation equilibrium and non-equilibrium ionisation thermal emission models ({\small XSPEC}
models \texttt{vapec} and \texttt{vpshock}, respectively), typical for shock-heated plasma in SNRs. We also attempted to fit the abundances of the main elements in the energy range of the SNR emission, namely O, Ne, Mg, and Fe. However, meaningful constraints were only obtained for the entire shell spectrum and not for the smaller regions due to limited statistics. For these spectra and all other elements, abundances were fixed to that of the LMC hot gas. Best-fit spectral parameters are listed in Table~\ref{tabsnrspec}. Improvements to the fit quality with the non-equilibrium ionisation model were only marginal. Regardless of ionisation equilibrium status, the overall absorption is low, indicating an object on the nearer side of the LMC gaseous disk. As expected from the X-ray images and dearth of emission at $E \gtrsim 1$~keV, the plasma temperature is relatively low, at $kT < 0.4$~keV. 

No strong spatial variations can be identified, all the parameters being consistent between eastern and western part of the shell, within the larger uncertainties stemming from the reduced statistics of the smaller regions. 
The emission from the superbubble region cannot be distinguished as well for the same reason as given above.
With both the \texttt{vapec} and \texttt{vpshock} models, the abundances of O, Ne, Mg, and Fe and their ratios are the same: They are those of the LMC hot gas phase \citep{2016A&A...585A.162M,2016AJ....151..161S} and reveal no enhancement by SN ejecta.

The total observed luminosity in the 0.3--8~keV band is $2.2 (\pm 0.06)\times 10^{35}$~erg~s$^{-1}$, lower than reported in \citet{2000AJ....119.2242C}, likely because we measured a slightly lower plasma temperature and higher $N_H$ (with better spectral resolution than \rosat) and excluded point sources previously unresolved. This sets \srce among the brightest third of LMC SNRs \citep{2016A&A...585A.162M}. However, given the large extent, the source is in the 20~\% \textit{faintest} SNRs when comparing surface brightness, which is the limiting factor for the study of extended sources.



\begin{figure}
\includegraphics[width=0.7\columnwidth,angle=-90]{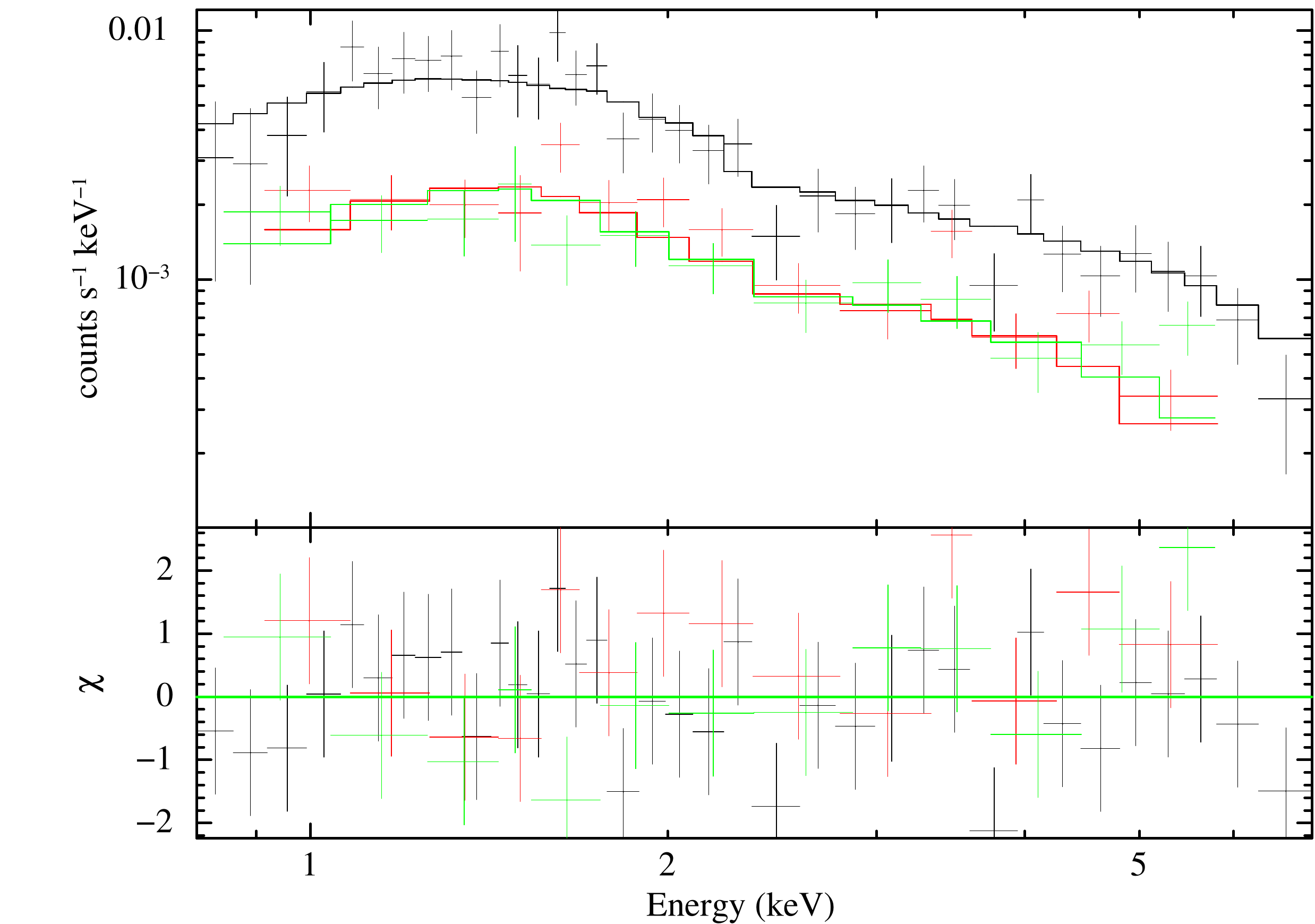}
\caption{EPIC spectra of the HMXB \src. Black, red and green denotes PN, MOS1 and MOS2 data points and model (histogram), respectively. The spectra have been rebinned for visual clarity.}
\label{figspechmxb}
\end{figure}

\subsubsection{HMXB}
  
As in the case of the X-ray timing analysis, only data from Obs. 2 were of sufficient statistical quality to perform a detailed spectral analysis.
The X-ray spectrum of the HMXB \src contains some contribution from the overlapping extended SNR as can be seen from Fig.~\ref{fig1}. In order to model this contribution, we included a component for the SNR emission in the spectral fit with the normalisation component left free --- since the surface brightness of the SNR is not uniform, we do not constrain the contamination level to the fractional geometric area covered by the source extraction region. The X-ray spectrum of the HMXB can be satisfactorily modelled with an absorbed power-law. The absorption scheme was the same as for the SNR, but the column density for the HMXB, $N_{H} ^{{\rm local}}$, represents both LMC gas in front of the source and potential local contributions.

The spectral parameters for the best-fit model using an absorbed power law are listed in Table~\ref{tabhmxbspec} and the spectra and best-fit model are shown in Fig.~\ref{figspechmxb}. The source was detected with an average absorption-corrected luminosity of 9$\times$\oergs{34}. Assuming that the source exhibited a similar spectrum during Obs.1 and from the PN count rate, we estimated an absorption corrected luminosity of 7$\times$\oergs{34} for \src during Obs. 1.

The local absorption component $N_{H} ^{{\rm local}}$ measured in the case of \src is significantly higher than that measured for the SNR (see Table~\ref{tabhmxbspec}). 
In order to estimate the total amount of absorption column density towards the direction of the source, we extracted the spectrum from the quasar MQS J050736.44-684751.6 which overlaps with the SNR along the line of sight (Fig.~\ref{fig1}). The spectrum was well-fitted by an absorbed power-law taking into account the determined redshift of the source. The best-fit parameters correspond to $\Gamma=2.3_{-0.3}^{+0.2}$ and N$_{\rm H}$=$3\times$ 10$^{21}$\,cm$^{-2}$. This is only slightly higher than the average integrated LMC column ($2.0\times 10^{21}$~cm$^{-2}$, Sect.~\ref{analysis_spectral_SNR}), perhaps due to further contribution by the host galaxy or intrinsic to the AGN. Nevertheless, this is several times lower than estimated from the HMXB \src which is very close to the quasar in projection. This establishes that the absorption of \src is not originating from the LMC large-scale structure. Instead, it indicates that $N_{H} ^{{\rm local}}$ is dominated by a local absorbing column density as is often detected in the case of HMXBs in the MCs \citep{2013A&A...558A..74V,2015MNRAS.447.2387C,2018MNRAS.480L.136M}. 

\begin{table}
\caption{X-ray spectral parameters of the HMXB \src for a power-law model.}
\begin{tabular}{lc}
 \hline
   Parameter & Value \\
 \hline
   $N_{H} ^{{\rm local}}$  (10$^{22}$ cm$^{-2}$) & 0.6$^{+0.4}_{-0.3}$ \\
   $\Gamma$ & 1.2$\pm0.2$ \\
   Flux $^a$ (0.2--12.0\,keV) &  2.6~$\pm$~0.3 \\
   Flux (unabsorbed) $^a$ (0.2--12.0\,keV) &  3.0~$\pm$~1.0 \\
   Absorption corrected X-ray luminosity$^{a}$ (erg~s$^{-1}$) & $8.5\pm1.0$~$\times~$$10^{34}$ \\
   \hline
    $\chi^{2}$       & 58.02  \\
   Degrees of Freedom & 59    \\
   \hline
\end{tabular}

$^{a}$Flux in units of 10$^{-13}$ \ergcm\ and assuming a distance of 50\,kpc in the energy band of 0.2--12\,keV\,. \\
$^{b}$ Absorption in units of $10^{22}~cm^{-2}$
Line-of-sight Galactic absorption was fixed to 6 $\times$ $10^{20}~cm^{-2}$\\
$^{c}$Errors are quoted at 90\% confidence.

\label{tabhmxbspec}
\end{table}

\section{Discussion}
\label{Sect:discussion}
We report here the positional coincidence of the BeXRB \src near the geometrical centre of \srce\, and the possible association of the two sources. We performed detailed timing and spectral analysis of the BeXRB using \xmm, \swift, eROSITA, and OGLE data and confirmed its nature as a highly magnetized neutron star. We also performed detailed X-ray spectral analysis of \srce using \xmm observations and compared with optical and radio observations to understand its nature.  

\subsection{Nature of \src: a BeXRB pulsar in the LMC}
Using the latest \xmm observation where the source was in the field of view (Obs.2) we discovered highly significant pulsations at 570\,s which confirm the nature of \src as a new BeXRB pulsar in the LMC. A total of $\sim$60 (candidate) HMXBs are known in the LMC out of which $\sim$90\% are Be/X-ray binaries and the rest are supergiant systems \citep{2021inpress,2020ATel13609....1H,2020ATel13610....1M,2019MNRAS.490.5494M,2018MNRAS.475.3253V,2018MNRAS.475..220V,2016MNRAS.459..528A}. Pulsations have been detected from only a small fraction of them (21) until now. This source significantly improves our knowledge of BeXRB pulsars in the LMC especially due to its possible association with SNR \srce which allows an estimate of its age. The X-ray intensity varies on long time scales as seen by comparing the \swift observations in Table~\ref{tabxrayswift}. The eRASS1 scans also displayed variability on shorter timescales.

The optical light curve of \src in the I-band spanning 13 years is highly variable in nature showing dip-like features as is typical for Be stars with binary companion. An intriguing fact is that the \xmm Obs. 2 and the \swift detection fall into `dips` in the optical emission as seen from the I-band light curve. The I vs. V--I colour-magnitude evolution displays that the emission becomes redder during the rise of the optical brightness, and asymptotically reaches a minimum value in I-V color. This indicates that as the Be disc grows in size, the optical emission gets brighter while the red continuum increases, a behaviour similar to what is seen in other Magellanic Cloud BeXRBs \citep[e.g.][]{2017A&A...598A..69H,2014A&A...567A.129V,2012MNRAS.424..282C}

\subsection{Nature and properties of \srce}
\label{discussion_SNR}
The shell-like morphology of \srce combined with its X-ray spectrum is typical of an middle-aged to old SNR. What sets it apart is its extreme diameter of 150~pc, easily the largest when compared to the population of confirmed LMC SNRs \citep[the largest SNR in][is 128~pc $\times$ 82~pc]{2017ApJS..230....2B}. Combined with the lack of obvious shell counterparts in optical and radio, the most natural explanation of the properties of \srce is that it is indeed a supernova remnant, but expanding into a very tenuous environment.

One can estimate properties such as ambient density, explosion energy, and age of the SNR based on its morphological and spectral parameters, under the assumption that it is in the Sedov phase \citep[e.g.][]{2004A&A...421.1031V,2012A&A...546A.109M}. The electronic density $n_e$ is calculated from the emission measure (Table~\ref{tabsnrspec}) and corresponding emitting volume $V$. For the latter we assume a spherical volume with a radius $R_{{\rm av}}$ averaged between semi-major and semi-minor axes.
It results in $n_e = 1.9 \, (\pm0.4) \times10^{-3} f^{-1}$~cm$^{-3}$ for the \texttt{vapec} model, and $n_e = 1.2 \, (\pm0.1) \times10^{-3} f^{-1}$~cm$^{-3}$ for the \texttt{vpshock} model, with $f$ the filling factor of the plasma within the volume ($f \leq 1$). Thus, it is clear that \srce is expanding in a rarefied medium.
Still, given the size of the shell, the total swept-up mass $M_{{\rm sw}} \propto n_e V$ within it is large, ranging from $881 \, (\pm 190)~f^2~M_{\odot}$ to $M_{{\rm sw}} = 557 \, (\pm54)~f^2~M_{\odot}$ for the \texttt{vapec} and \texttt{vpshock} models, respectively. Even accounting for a small filling factor $f$, $M_{{\rm sw}}$ is much in excess of the ejecta and circumstellar material mass for any type of progenitor, and is therefore dominated by ambient ISM, justifying the assumption of a Sedov phase.
The SN explosion energy is set by the size of the shell $R$, the plasma temperature $k T_s$ (which depends on the shock speed), and the ambient density as $E_0 \propto k T_s R_{{\rm av}}^3 n_e$. We found $E_0 = 1.0 \, (\pm0.3)\times 10^{51} f^{-1}$~erg and $E_0 = 0.9 \, (\pm0.2)\times 10^{51} f^{-1}$~erg for the \texttt{vapec} and \texttt{vpshock} models, respectively. This is close to the canonical $10^{51}$~erg for a SN and thus consistent with the interpretation of the energetics of \srce being dominated by a single SN.
Finally, the dynamical age of \srce $t_{{\rm dyn}} \propto R_{{\rm av}} (kT_s)^{-1/2}$ is in the range 55 to 63~kyr in the collisional ionisation equilibrium case, and between 43 and 54~kyr for the non-equilibrium case, as the temperature is slightly higher. \srce is one of the oldest LMC SNRs, as expected at such a large size. The derived ambient density, explosion energy and age of the SNR are consistent with the estimates of \citet{2000AJ....119.2242C}.

Given the large size, low density, and complete shell morphology, \citet{2000AJ....119.2242C} suggested that the SNR was located in the near side of the LMC halo. Here we suggest another possibility. The NGC~1850 cluster, which is located at the same position is a double cluster comprising a large globular-like cluster separated from a smaller, younger compact cluster, known as NGC~1850B, by $\sim30\arcsec$ \citep[e.g.][]{Robertson1974,Fischer1993, Gilmozzi1994}. The ages of both clusters have been determined in various studies. \citet{Gilmozzi1994} determined ages of $50\pm10$~Myr and $4.3\pm0.9$~Myr for the main cluster and NGC~1850B, respectively. Other studies are in general agreement, with estimates ranging from 40--100~Myr for the main cluster and 1--10~Myr for NGC~1850B \citep[e.g.][]{Fischer1993,Vallenari1994}. Mass estimates for NGC~1850 are  
$\sim10^{4-5}$~M$_{\sun}$ \citep{Fischer1993, McLaughlin2005}, placing it among the most massive in the LMC outside 30~Dor. Given the age of the main NGC~1850 cluster, the massive stellar population has already been lost to SNe, and there are no stars remaining that are capable of photoionising the N103 shell. Rather, the young massive stars of NGC~1850B are responsible \citep{Fischer1993,Ambrocio1997}. However, in its infancy, the main NGC~1850 cluster would have been a powerhouse, containing a significant massive stellar population, more than capable of driving a large superbubble into the ISM. To explain the low ambient density inferred for the SNR, we propose the following scenario: \textit{(i)} the main NGC~1850 cluster created and powered a superbubble in the region; \textit{(ii)} the evolution of this superbubble stalled after $\sim40$~Myr when the massive stellar population was lost and the internal pressure dropped below the ISM pressure, leaving the large, low density superbubble relic; \textit{(iii)} NGC~1850B formed near the main cluster and its massive stars are now photoionising part of the original superbubble shell, which we know as N103; \textit{(iv)} the SNR progenitor exploded in the interior of the superbubble relic created by the NGC~1850 main cluster. We also note that NGC~1850 and N103 are projected against the inside edge of the supergiaint shell SGS5 \citep{Kim1999} and the initial superbubble could have blown out into this region. In any case, the proposed scenario can explain the very low ambient density into which the SNR has evolved.

\subsection{A new BeXRB-SNR association ?}
Using \rosat observations, \citet{2000AJ....119.2242C} identified a compact source at the center of \srce superposed on the star cluster HS122. This was proposed to be a BeXRB based on its coincidence with a star-cluster region, although no timing or spectral analysis could be performed to understand its nature. The position of the compact source proposed in \citet{2000AJ....119.2242C} is compatible with the now known quasar MQS J050736.44-684751.6, which overlaps with the SNR along the line of sight. The other point source in the vicinity, \src has been identified as a BeXRB pulsar instead in this work which qualifies as the most probable compact object associated with \srce. 
It is likely that \src, owing to its intrinsic X-ray variability, was in a faint state during the \rosat observations presented in \citet{2000AJ....119.2242C}, and thus either undetected or confused with the quasar.

For the BeXRB to be associated to the SNR requires a core-collapse origin, with the parent SN producing both the remnant and leaving an NS behind. \srce is in a star-forming region as evidenced from the presence of many upper main-sequence stars in its surrounding, and confirmed by the reconstructed star-formation history by \citet{2009AJ....138.1243H}. In a galaxy seen almost face-on like the LMC, the location of an SNR in such a star-forming region is a strong indication of a core-collapse origin, as opposed to galaxies with significant line-of-sight depth like the SMC where more confusion can arise. In the LMC, only one type Ia SNR can be clearly misidentified by looking solely at the star-forming environment of the object \citep[][]{2016A&A...585A.162M}. Coincidentally, this source is the nearby SNR N103B. Even in the absence of strong spectral evidence for a core-collapse SN, because the emission is dominated by LMC ISM, we thus conclude that \srce was most likely created in a core-collapse SN.

\begin{figure}
    \centering
    \includegraphics[width=1.0\columnwidth]{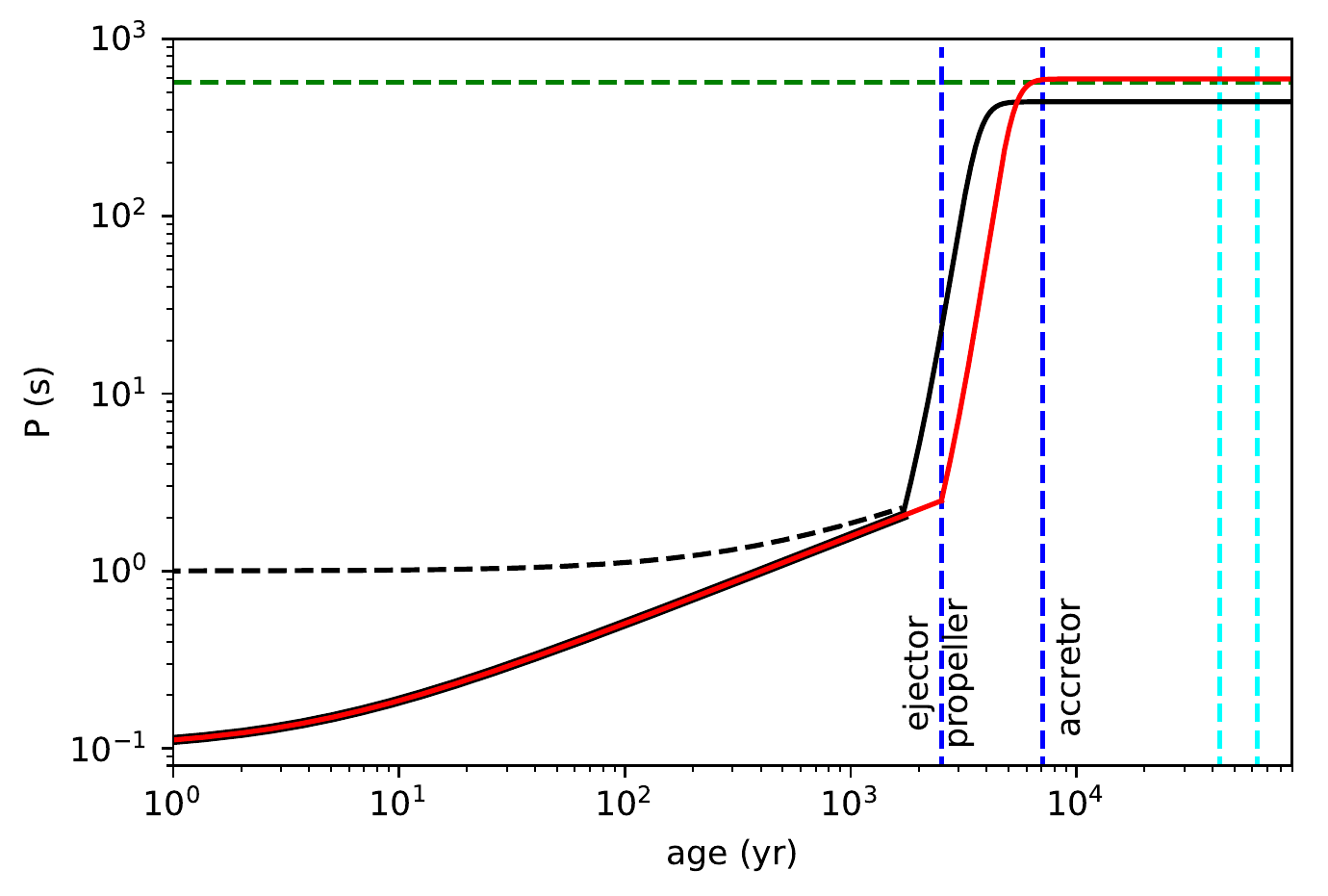}
    \caption{Spin evolution as a function of time marking the ejector, propeller and ejector phases of \src following \citet{2020MNRAS.494...44H}.
    Assumed initial spin of the neutron star: $P_{\rm 0}$ is 100 ms (solid lines) and 1\,s (dashed line). $B=4\times10^{14}$\,G and evolution is shown for a constant mass accretion rate of $\dot{M}=8\times10^{-12}$~$M_{\odot}$\,yr$^{-1}$ (red) and $2\times10^{-11}$~$M_{\odot}$\,yr$^{-1}$ (black).
    The green horizontal line marks the spin period and the cyan lines indicate the estimated age of the system, assuming the pulsar is indeed the progenitor of \srce.}
    \label{fig:spin_evo}
\end{figure}

Furthermore, the proximity of \src near the geometrical centre of the SNR (0.9\arcmin), indicates the association of the BeXRB pulsar \src with \srce. As the SNR 
has an almost perfect elliptical shape, the center of the ellipse is a good proxy for the likely explosion site and the origin of the NS in \src. 
This sets constraints on the NS transverse velocity. At the LMC distance of 50~kpc, the transverse velocity projected on the sky is: 
\begin{equation}
    \label{eq_kickvelocity}
    v_{{\rm proj}} = 284 \ \delta \theta \ \left(\frac{D}{50\ {\rm kpc}}\right) \left(\frac{t_{{\rm SNR}}}{{\rm50\ kyr}}\right)^{-1} \ {\rm km\ s}^{-1}
\end{equation}
with $\delta \theta$ the angular separation in minutes of arc from the geometric centre of the SNR to the NS, and $t_{{\rm SNR}}$ the age of the SNR. Equating this to the dynamical age $t_{{\rm dyn}}$  estimated for a Sedov model (Sect.~\ref{discussion_SNR}, ranging from 43 to 63~kyr, and for $\delta \theta =0 .9$\arcmin, the projected velocity is 200 to 300~km~s$^{-1}$. The observed velocity distribution of radio pulsars
indicate a mean pulsar transverse velocity of 345~km~s$^{-1}$, with an inferred three-dimensional velocity of 450~km~s$^{-1}$ \citep{1994Natur.369..127L}. The prediction for NSs in HMXBs is expected to be smaller and has diverse values in literature \citep[on the order of 30--150\,kms$^{-1}$; see][]{2005MNRAS.358.1379C,2012ApJ...744..108B,2014MNRAS.437.1187Z,2015A&A...573A..58Z}.
The inferred transverse velocity of the BeXRB pulsar \src is slightly higher than expected from the current predictions and therefore further raises a question on its association with \srce. Moreover, it cannot be ruled out that even if the SNR and the NS originated from the same star cluster, the progenitors of both could be distinct. 

\begin{figure}
    \centering
    \includegraphics[width=1.0\columnwidth]{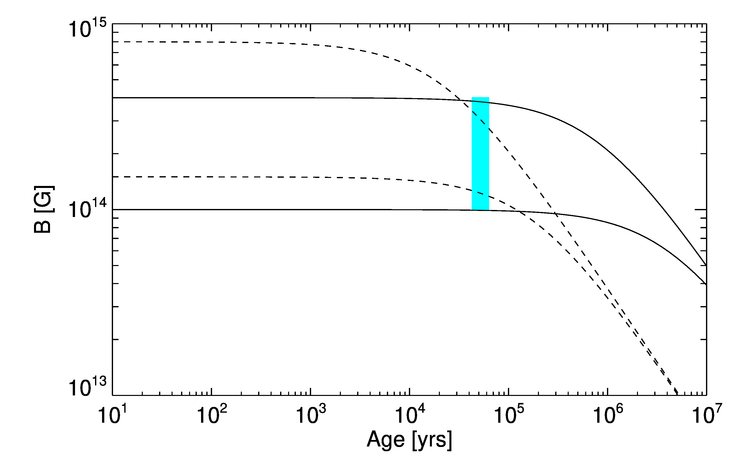}
    \caption{Magnetic field evolution assuming ambipolar diffusion in the irrotational mode (i.e. solid lines; $b$=0.01 and $\alpha$=5/4), or in the solenoidal mode (i.e. dashed lines; $b$=0.15 and $\alpha$=5/4), for different initial $B$ field strength. The loci of the calculated $B$ field assuming spin equilibrium for the age of the NS is marked with cyan color.}
    \label{fig:decay}
\end{figure}
\subsection{Constraints of NS properties at birth and the spin evolution of the system}




Evolutionary, NSs are thought to be born with short spin periods and strong magnetic fields 
\citep[$P_0\sim10-100$\,ms, $B\sim10^{13}$\,G;][]{2006ApJ...643..332F}. 
At birth the BeXRB pulsar is thought to be at the ejector phase. In this regime, the pulsar does not interact with the accreting matter as the NS light cylinder is smaller than the magnetospheric radius as would be calculated by the balance of magnetic and gas pressure. 
As the NS spins down, at some point the magnetospheric radius and light cylinder become equal and the system transitions to the propeller phase, where the interactions of the disk and magnetosphere cause a faster spin down rate. 
Eventually an equilibrium is reached, however the whole evolution is complicated by a time variable accretion rate. The duration of each phase is a function of the NS $B$ field, and average mass accretion rate throughout the evolution. Following the prescription of \citet{2020MNRAS.494...44H} we traced the evolutionary history of \src given the estimate of the magnetic field obtained by equating the co-rotation and magnetospheric radius at spin-equilibrium:
\begin{equation}
P_{\rm eq} = \left(\frac{4\pi^{2}R_{\rm M}^{3}}{GM_{\rm NS}}\right)^{1/2}, \,R_{\rm M} = \frac{\xi}{2} \left(\frac{R_{\rm NS}^{12}B^4}{GM_{\rm NS}\dot{M}^2}\right)^{1/7}
\label{eq:RM}
\end{equation}
where $G$ is the gravitational constant, $M_{\rm NS}=1.4$ M$_{\odot}$ and $R_{\rm NS}=10$~km is the NS mass and radius. This results into a magnetic field of $B\sim4\times10^{14}$\,G. Further, from the observed $L_{X}$ (i.e. $\sim$\oergs{35}) and by assuming that all dynamic energy of the in-falling material is converted to radiation we derive a mass accretion rate of  5$\times$10$^{14}$ g~s$^{-1}$ (i.e. $L_{X}{\approx}0.2\dot{M}c^2$). Using eq.7 and eq.12 from \citet{2020MNRAS.494...44H}, Fig.~\ref{fig:spin_evo} plots the complete evolution of the NS spin period from the ejector phase, to the propeller phase and finally to the accretor/spin equilibrium phase when the current measured spin period of the NS is achieved. From the measured $B$ and mass-accretion rate the system crosses the propeller stage at $t\simeq2600$\,yr and reaches spin-equilibrium at  $t\simeq7000$\,yr.
As is expected, the choice of the initial spin period $P_{0}$ does not affect the evolution of the system at later times, such as the accretor phase. We can obtain an upper limit on the birth spin of the NS by equating  $R_{\rm M}$ with $R_{LC}$ (light-cylinder) and find $P_0\lesssim2$ s.

The estimated magnetic field strength of $B\sim4\times10^{14}$\,G is obtained by a simple prescription of equating the co-rotation and magnetospheric radius and assuming the scaling factor between magnetospheric radius ($R_{\rm M}$) and Alfvén radius; $\xi\simeq0.5$ \citep{2018A&A...610A..46C}. The $B$ here also refers to the field measured at the magnetic poles ($\mu=BR^{3}/2$, where $\mu$ is the magnetic dipole moment) and the field strength measured at the magnetic equator is lower by a factor of 2. The results overall indicate a magnetar-like field strength of $B\gtrsim10^{14}$\,G.

The prescription of \citet{2020MNRAS.494...44H} assumes a constant mass accretion rate and $B$ throughout the various stages undergone by the system. The $B$ on the other hand decays through non linear processes such as ambipolar diffusion and/or a Hall cascade \citep{1992ApJ...395..250G}. Numerical simulations has shown that for $B>10^{13}$~G this evolution follows a simple decay law \citep{2000ApJ...529L..29C}, with the following form:
\begin{equation}
B(t)=\frac{B_{0}}{[1+b{\alpha}B_{0}^{\alpha}t]^{1/\alpha}},
\label{eq:B}
\end{equation}
where $B$ and $t$ quantities are in units of $10^{13}$~G and $10^6$ yr respectively, while values for parameters $b$ and $\alpha$ 
denote different modes of $B$ field decay \citep[see][for details]{2000ApJ...529L..29C}.

%
By following eq. \ref{eq:B} we can calculate the B field decay from its birth and compare it with its current value (see Fig \ref{fig:decay}). It is evident that minimal decay has occurred within this time and the pulsar has maintained its initial $B$ strength as is expected from the time scales expected from the magnetic field decay with is $\gtrsim10^{6}$ yr.

\section{Conclusions}
\label{Sect:conclusion}
The probable association of \src with \srce makes it the third identified BeXRB-SNR system. Pulsations are discovered at 570\,s indicating the spin period of the neutron star. The estimated age of the SNR is 43--63\,kyr which points to a middle aged to old SNR. The magnetic field strength {\bf of the neutron star} is indicated to be $\gtrsim10^{14}$\,G. The other 2 such systems are SXP\,1062 \citep{2012A&A...537L...1H,2012MNRAS.420L..13H} and SXP\,1323 \citep{2019MNRAS.485L...6G} both located in the Small Magellanic Cloud. This highlights the ideal environment and the suitability of finding such objects in the Magellanic Clouds. All of the three BeXRB-SNR systems (including our discovery) host slowly spinning pulsars and have similar estimates for the age of the parent SNR. The eROSITA instrument on board the Russian/German Spektrum-Roentgen-Gamma(SRG) mission all-sky survey will provide a deep and complete coverage of the Magellanic System for the first time in X-rays and will be instrumental in finding more such systems.


\section*{Acknowledgements}
This work uses observations obtained with \xmm, an ESA science mission with instruments and contributions directly funded by ESA Member States and NASA. The \xmm project is supported by the DLR and the Max Planck Society. 
GV is supported by NASA Grant Number  80NSSC20K0803, in response to XMM-Newton AO-18 Guest Observer Program. GV acknowledges support by NASA Grants number 80NSSC20K1107 and 80NSSC21K0213. M.S.\ acknowledges support by the Deutsche Forschungsgemeinschaft through the Heisenberg professor grants SA 2131/5-1 and 12-1.
The OGLE project has received funding from the National Science Centre,
Poland, grant MAESTRO 2014/14/A/ST9/00121 to AU.
This work is based on data from eROSITA, the primary instrument aboard SRG, a joint Russian-German science mission supported by the Russian Space Agency (Roskosmos), in the interests of the Russian Academy of Sciences represented by its Space Research Institute (IKI), and the Deutsches Zentrum f{\"u}r Luft- und Raumfahrt (DLR). The SRG spacecraft was built by Lavochkin Association (NPOL) and its subcontractors, and is operated by NPOL with support from the Max Planck Institute for Extraterrestrial Physics (MPE).
The development and construction of the eROSITA X-ray instrument was led by MPE, with contributions from the Dr. Karl Remeis Observatory Bamberg \& ECAP (FAU Erlangen-N{\"u}rnberg), the University of Hamburg Observatory, the Leibniz Institute for Astrophysics Potsdam (AIP), and the Institute for Astronomy and Astrophysics of the University of T{\"u}bingen, with the support of DLR and the Max Planck Society. The Argelander Institute for Astronomy of the University of Bonn and the Ludwig Maximilians Universit{\"a}t Munich also participated in the science preparation for eROSITA.
The eROSITA data shown here were processed using the eSASS/NRTA software system developed by the German eROSITA consortium.
The Australian SKA Pathfinder is part of the Australia Telescope National Facility which is managed by CSIRO. Operation of ASKAP is funded by the Australian Government with support from the National Collaborative Research Infrastructure Strategy. ASKAP uses the resources of the Pawsey Supercomputing Centre. Establishment of ASKAP, the Murchison Radio-astronomy Observatory and the Pawsey Supercomputing Centre are initiatives of the Australian Government, with support from the Government of Western Australia and the Science and Industry Endowment Fund. We acknowledge the Wajarri Yamatji people as the traditional owners of the Observatory site.
\section*{Data Availability}
X-ray data are available through the High Energy Astrophysics Science Archive Research Center \url{heasarc.gsfc.nasa.gov}. 
OGLE data are available through the OGLE XROM online portal \url{http://ogle.astrouw.edu.pl/ogle4/xrom/xrom.html}.


\bibliographystyle{mnras}
\bibliography{example} 






\bsp	
\label{lastpage}

\end{document}